# Space Environments Variability and its Impact on Total Dose and Single Event Effects in Electronic Parts
*Simulation and Modeling*

G.I. Zebrev[1], I. V. Elushov[2]

[1]*Department of Micro- and Nanoelectronics of National Research Nuclear University "MEPHI", Moscow, Russia,*
*gizebrev@mephi.ru*
[2]*JSC Scientific Research Institute of Space Instrument Engineering, Moscow, Russia*

Abstract: The aim of this paper is the modeling and simulation of impact of space radiation variability on total dose and single event effects in spaceborne electronics. It has been shown that the simultaneous thermal annealing may lead to non-stationary relaxation after dose-rate peaks. Significant enhancement of soft error rate during solar flares in the memories mitigated by the scrubbing due to non-linear dependence on particle flux has been revealed.

## 1 INTRODUCTION

Most testing experiments on the total ionizing dose (TID) and the single event effects (SEE) have been typically performed with a constant dose rate or with a fixed particle flux. In reality the space environments are highly variable, and the dose rates and particle fluxes are typically orders of magnitude larger during solar flares (see, e.g. (Barth, 2003; Turflinger, 2004). The impact of the space radiation environment variability on spacecrafts can affect both the total dose and single event effects in spaceborne microelectronic parts.

Space weather variability makes predictions, for example, of the SEE difficult, in particular, because the future circuits are likely to have more complex radiation response due to the use of the internal mitigation systems such as scrubbing.

We anticipate also that simultaneous time-dependent anneal could significantly modify radiation response of space-borne integrated circuits under the solar flares against the long-term low-dose rate radiation background.

The objective of this report is to improve knowledge of space radiation variability impact via development of modeling approaches to predict the circuit radiation response.

## 2 REAL-TIME DOSE RATE MONITORING SYSTEM

The Russian Federal Space Agency Monitoring System of space radiation environments includes two parts: the scientific monitoring system (ground-based segment) and the engineering monitoring system (space-borne segment). The space-borne segment contains a set of the small-size n-MOSFET based dosimeters connected with telemetric sub-systems for permanent data transfer to the Earth.

The space-borne TID dosimeters provide (Anashin, 2010) (Anashin, 2011)
- output electrical signal is proportional to TID;
- real-time monitoring of TID;

They have the voltage and current control for providing a better linearity and temperature stabilization by operating-point selection according to minimal change of sensor current-voltage curve. The main characteristics of dosimeters are
- the proportional response in the output range 0.1-100 krad;
- frequency output – 1-200 kHz;
- RS-485 interface.

The dosimeters calibration was carried out at the facility "GU-200" with $^{60}$Co isotopes (Research Institute of Scientific Instruments, RISI, Lytkarino, Moscow district). The flight data at the circular orbits ~20000 km has been obtained from October 2008.

# 3 VARIABILITY OF DOSE RATES IN SPACE

## 3.1 Rate equation approach in modeling

A common cause of the parameter degradation in the CMOS and the bipolar technology parts is the radiation-induced defect buildup in the insulation layers and at the interfaces. The time dependent thermal annealing of these defects has found to be a universal effect, and the devices of all technologies to some extent are prone to it, although the annealing parameters (activation energies and the time constants) are specific for different technologies, or even for different lots of devices.

The kinetic equation for the changes of a criterion parameter $\Delta\Pi$ can be written in the linear response approximation as

$$\frac{d\Delta\Pi(t)}{dt} = A\eta_{eff}\left[P(t), T_{irr}(T), E_{ox}(t)\right]P(t) - \frac{\Delta\Pi(t)}{\tau_a\left[T_{irr}(t)\right]}, \quad (1)$$

where $A$ is a technology and lot dependent normalized factor, $\eta_{eff}$ is the effective charge yield function, generally dependent on the electric field in the oxides $E_{ox}$, irradiation temperature $T_{irr}$, and a dose rate $P$. The simultaneous thermal annealing is described in (1) with the anneal temporal constant of a single (in the simplest case) or a few (or, even, continuum energy spectrum) of defects.

Rate equation (1) has a general solution for arbitrary profile of dose rate $P(t)$

$$\Delta\Pi(t) = A\int_0^t \eta_{eff}\left[P(t'), T_{irr}(t'), E_{ox}(t')\right]e^{-\frac{t'-t}{\tau_a\left[T_{irr}(t')\right]}}P(t')dt'$$

This solution explicitly demonstrates the fact that the radiation degradation is a functional (a convolution) of the previous dose-rate, thermal and electric history of the exposed object. This means that any non-stationary variation in dose-rate environments, electric and thermal conditions would lead to time-dependent transient effects in parameters degraded under irradiation.

## 3.2 Space Dose-Rate Variations

An increase in $\eta_{eff}$ with decreasing dose rate in the bipolar ICs is known as the ELDRS effect (Pease, 2009). The explicit form of this function was found in (Zebrev, 2006, 2009). According to (Zebrev, 2006) the enhanced charge yield at low dose rate is possible due to a decrease in electron-hole recombination via the Langevin mechanism through steady-state and dose-rate dependent density of localized holes. This recombination is suppressed by the strong electric fields in the insulating oxides and therefore not observed in the gate oxides of the MOSFETs (except, may be, the thick oxides of p-MOSFET based dosimeters (Kim, 2003).

It is worth to notice that even large dose-rate temporal fluctuations do not change significantly the charge yield in the range $10^{-6}$-$10^{-3}$ rad/s, where it only weakly depends on $P(t)$ exhibiting the magnitudes nearby the maximum at a given temperature and electric field in the oxide. This means that the ELDRS effect is rather a problem of the ground test adequacy than a problem of space.

## 3.3. Time-Dependent Simultaneous Anneal

In bipolar circuits the moderate elevation of irradiation temperatures (up to 100 °C) for short-term exposure (when the time-dependent anneal did not have time to become noticeable) enhances the degradation due to the increase of the charge yield $\eta_{eff}$ (Zebrev, 2006). However, under the long-term irradiation the simultaneous thermal anneal limits the degradation in all types of devices. Competition between enhancement of the charge yield and the simultaneous thermal annealing results in an appearance of the remarkable degradation maximum in bipolar transistors as function of irradiation temperature (Witczak, 1996). Thereby taking into account the thermal annealing in prediction procedure would lead to a noticeable reduction in the conservatism of the hardness assessment.
Particularly, the simultaneous thermal annealing may lead to the time-dependent phenomena of the degradation relaxation after the dose-rate peaks. As an illustration, Fig. 1 shows the simulation results for the bipolar IC degradation calculated for the actual dose-rate profile derived from the on-board satellite date (Barth, 2003) (Turflinger, 2004).

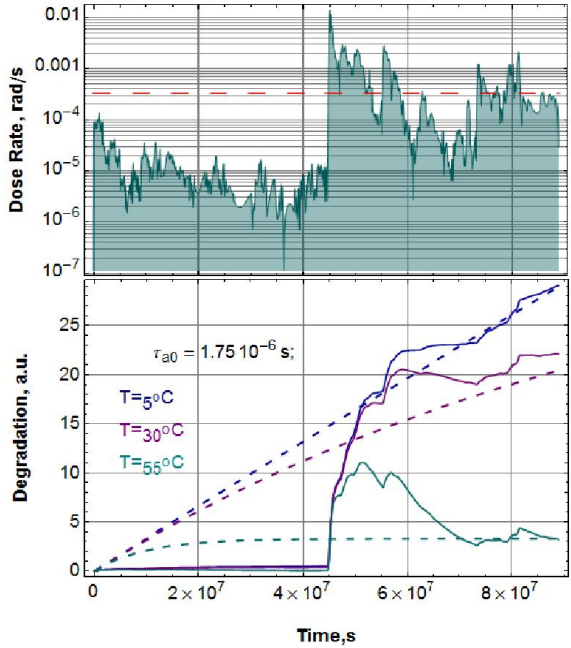

Figure 1: Measured dose rate temporal profile on the satellite orbits ~20000 km with a remarkable peak in April 2010 (top figure) and simulated degradation curve (in arbitrary units) calculated for three irradiation temperatures with the temporal annealing constants obtained in [7] (bottom figure), based on experimental data at different operation temperatures [9].

At the same time, it is important, that elevated operation temperatures may invoke an occurrence of quasi-steady-state saturation of degradation due to dynamical compensation of the radiation-induced buildup and the simultaneous annealing processes. Specifically, the level of the degradation saturation generally depends on dose rate, irradiation temperature and electric regime, implying that non-stationary and non-monotonic effects could be observed after variation in conditions.
Quasi-steady-state saturation in degradation implies also an opportunity of the sensitivity to the operation temperature variations, and, particularly, the anti-correlation between the instant operation temperature and degradation effect magnitude.

## 4 SOFT ERROR RATE UNDER SCRUBBING

### 4.1 Soft Error Rate in Complicated Systems with Scrubbing

The instant soft error rate in the simple SRAM matrices is characterized by the number of the events per unit time $\lambda_0(t)$ which can be represented as a product of the effective cross-section $\sigma_0$ by the instant particle flux value $\phi(t)$:

$$\lambda_0(t) = \sigma_0 \phi(t). \qquad (3)$$

Here, for the sake of simplicity and brevity we have deliberately oversimplified the description of SER assuming that $\sigma_0$ and $\phi(t)$ are the SEU saturation cross-section and a flux particles having the LETs (or, for proton, energy) above its threshold values (the unit-step approximation).

For simple memory matrices the number of errors $N^{err}(t)$ does not depend on the particle flux temporal profile and is proportional to the fluence $\Phi(t)$

$$N^{err}(t) = \sigma_0 \int_0^t \phi(t')dt' = \sigma_0 \Phi(t). \qquad (4)$$

At the same time, the present-day ICs, e.g. the FPGA, have typically more complicated structure and often possess the built-in self-repairing mechanisms known as scrubbing (Mukherjee, 2008), which imply periodical inspection and data correction in memory cells. Such systems have more complicated reaction to non-stationary fluxes of particles. Let us consider a memory array with $N_W$ words having $M$ memory cells in each line (see Fig. 2).

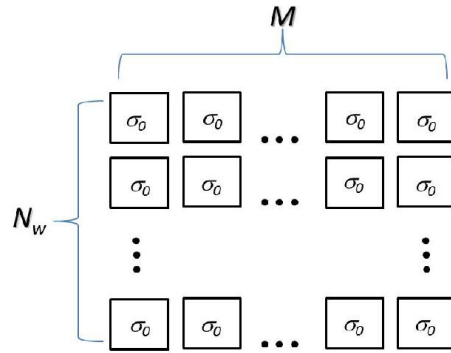

Figure 2: Memory cell matrix.

Scrubbing procedure corrects a word after each inspection operation with a characteristic time $t_R$. The correction provides the following failure condition: an error occurs only when not less than 2 upsets take place in the same word per a time $t_R$. The event density distribution function for the Poisson processes $f(t)$ is expressed as follows:

$$f(t) = \lambda \exp(-\lambda t). \quad (5)$$

The effective rates for the first $\lambda_1$ and for the second $\lambda_2$ upset in the word are determined

$$\lambda_1 = M\sigma_0\phi, \quad \lambda_2 = (M-1)\sigma_0\phi. \quad (6)$$

The distribution function $f_d(t)$ for the two successive upsets per a time $\Delta t$ in the same word can be described as the distribution's product integrated over all possible intermediate times

$$f_d(\Delta t) = \int_0^{\Delta t} f_2(\Delta t - t')f_1(t')dt' = \frac{\lambda_1\lambda_2}{\lambda_1-\lambda_2}\left(e^{-\lambda_2\Delta t} - e^{-\lambda_1\Delta t}\right) \quad (7)$$

This is nothing but distribution density of errors in the words with scrubbing. At small flux and/or times $\lambda\Delta t \ll 1$ the Taylor series expansion of (7) yields

$$f_d(\Delta t) \cong \lambda_1\lambda_2\Delta t = M(M-1)(\sigma_0\phi)^2 \Delta t. \quad (8)$$

The error number $\Delta N^{err}$ in M words per a scrubbing time $t_R$ is proportional to the number of words and cumulative upset probability

$$\Delta N^{err}(t_R) = N_W \int_0^{t_R} f_d(\Delta t) d\Delta t$$
$$= \frac{N_W}{2} M(M-1)(\sigma_0\phi t_R)^2. \quad (9)$$

In the course of a space mission $t_{mission}$ or during the ground tests the number of errors will be proportional to the number of scrubbing cycles:

$$N^{err}(t_{mission}) = \Delta N^{err}(t_R)\frac{t_{mission}}{t_R} =$$
$$= \frac{N_W}{2} M(M-1)(\sigma_0\phi)^2 t_R t_{mission}. \quad (10)$$

Thus, the mean SER for long-term missions ($t_{mission} \gg t_R$) turns out to be proportional to the scrubbing time. Using (10) one can obtain the mean cross-section value per a word (not per a cell!) taking into account the scrubbing mitigation

$$\sigma_W(t) = \frac{N^{err}(t_{mission})}{N_W \phi t_{mission}} = \frac{1}{2} M(M-1)(\sigma_0\phi(t)t_R)\sigma_0. \quad (11)$$

Note that the effective error cross-section for the systems mitigated by scrubbing turns out to be proportional to the instant flux value. An increase in the scrubbing frequency reduces the upset probability, but takes a part of computational resources and power consumption of the system. Characteristic time $t_R$ can be chosen as a trade-off between minimization of error rate and consumption.

Using the flux value averaged over the scrubbing cycle one can derive that the number of upsets in time period $t_{mission} \gg t_R$ is determined by an integral of squared flux and not by total fluence as in a simple memory array case (1)

$$N^{err}(t_{mission}) = \frac{N_W}{2} M(M-1)\sigma_0^2 t_R \int_0^{t_{mission}} \phi^2(t) dt. \quad (12)$$

This fact is very important because the proton fluxes in space have fluctuations as much as several orders of magnitude (see Fig. 3 with GOES satellite data taken for a random time period (http://www.swpc.noaa.gov/ftpmenu/lists/particle.html).

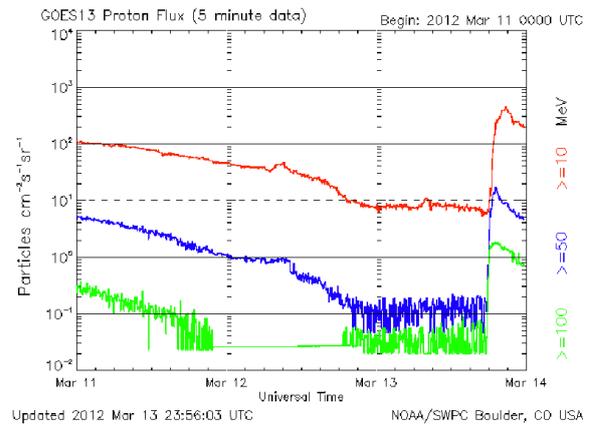

Figure 3: GOES flux data for protons 10, 50, and 100 MeV (11-14 March 2012).

Great temporal dispersion of particle fluxes during the space mission may result in the fact that the error number per a given time would depend not so much on the total fluence as on the time profile of particle flux during this time. Particularly, at the same fluence the ICs mitigated by scrubbing may parry the most part of the errors under a uniform flux while it may demonstrate much greater vulnerability to soft errors because of the scrubbing procedures were overloaded at the moments of sharp particle flux peaks.

# CONCLUSION

- A real degradation in space is a functional of the radiation, electric and thermal history of the devices on the board of the spacecraft;
- Neglecting of temporal variability of space particle flux may lead to significant underestimation of soft error rate in circuit mitigated by scrubbing.

# REFERENCES

Anashin V.S., Ishutin I.O., et al. 2010. "Exploitation of space ionizing radiation monitoring system in Russian Federal Space Agency," 7 European Space Weather Week.

Anashin V.S., Protopopov G.A., et al. 2011. "Monitoring of Space Ionizing Radiation in Russian Federal Space Agency", 8 European Space Weather Week, 2011.

Barth, J., et al. 2003. "Space, Atmospheric, and Terrestrial Radiation Environments," IEEE Trans. on Nucl. Sci., vol. 50, no. 3, pp. 466-482,

Kim S J. et al., 2003. "Enhanced Low Dose Rate Sensitivity (ELDRS) Observed in RADFET Sensor", RADECS 2003 Proceedings, pp.669-671.

Mukherjee Shubu, 2008, chap.5 in *Architecture Design For Soft Errors*, Morgan Kaufmann Publishers.

Pease R. L., Schrimpf R. D. and Fleetwood D. M., 2009. "ELDRS in Bipolar Linear Circuits: A Review," IEEE Trans. on Nucl. Sci., Vol. 56, No. 4, pp. 1894-1908.

Turflinger T., et al., 2004. "Solar Variability and the Near-Earth Environment - Mining ELDRS Data From the Microelectronics and Photonics Test Bed Space Experiment," NASA/TP - 2004-213339

Witczak S. C., Schrimpf R. D., Galloway K. F., Fleetwood D. M., Pease R. L., Puhl J. M., Schmidt D. M., Combs W. E., and Suehle J. S., 1996. "Accelerated tests for simulating low dose rate gain degradation of lateral and substrate PNP bipolar junction transistors," IEEE Trans. Nucl. Sci., vol. 43, pp. 3151–3160.

Zebrev G.I. et al., 2006. "Radiation Response of Bipolar Transistors at Various Irradiation Temperatures and Electric Biases: Modeling and Experiment", IEEE Trans. on Nuclear Science, Vol. 53, No. 4, pp. 1981-1987.

Zebrev G.I., Gorbunov M.S., 2009. "Modeling of Radiation-Induced Leakage and Low Dose-Rate Effects in Thick Edge Isolation of Modern MOSFETs" IEEE Trans. on Nuclear Science, Vol. 56, No. 4, pp. 2230-2236.